\documentclass{revtex4}
\usepackage{amssymb}
\usepackage{amsfonts}
\usepackage{amsmath}
\usepackage{graphicx}
\usepackage[font={footnotesize,it}]{caption}

\begin{document}

\preprint{}
\title{Black p-Branes versus black holes in non-asymptotically flat
Einstein-Yang-Mills theory}
\author{S. Habib Mazharimousavi}
\email{habib.mazhari@emu.edu.tr}
\author{M. Halilsoy}
\email{mustafa.halilsoy@emu.edu.tr}
\affiliation{Department of Physics, Eastern Mediterranean University, G. Magusa, north
Cyprus, Mersin 10 - Turkey.}
\date{\today }

\begin{abstract}
We present a class of non-asymptotically flat (NAF) charged black p-branes
(BpB) with p-compact dimensions in higher dimensional Einstein-Yang-Mills
theory. Asymptotically the NAF structure manifests itself as an
anti-de-sitter spacetime. We determine the total mass / energy enclosed in a
thin-shell located outside the event horizon. By comparing the entropies of
BpB with those of black holes in same dimensions we derive transition
criteria between the two types of black objects. Given certain conditions
satisfied our analysis shows that BpB can be considered excited states of
black holes. An event horizon $r_{+}$ versus charge square $Q^{2}$ plot \
for the BpB reveals such a transition where $r_{+}$ is related to the
horizon radius $r_{h}$ of the black hole (BH) both with the common charge $%
Q. $
\end{abstract}

\pacs{PACS number}
\maketitle

\section{Introduction}

It is well-known that by uplifting $d-$dimensional dilatonic black holes
(BH) one obtains $\left( d+p\right) -$dimensional black $p-$branes (BpB)
with extended event horizons \cite{1,2,3,4}. A black string (BS), for $p=1$
is an extension with one extra dimension whose horizon has a product
topology such as $R\times S^{d-2}$. Naturally the simplest member of this
class constitutes the chargeless Schwarzschild metric in $d=4$ and $p=1$, so
that the topology becomes $R\times S^{2}$ \cite{5,6}. In case that the extra
dimension is compact the end points may be identified to give a BS with
horizon topology $S^{1}\times S^{2}$. Concerning BS (and more generally BpB)
an interesting problem that gave birth to a considerable literature in
recent years is their instability against decay into BH (or vice versa).
This problem was pointed out first, through perturbation analysis by Gregory
and Laflamme (GL) which came to be known as GL instability \cite{7,8}. Such
a perturbative stability / instability applies to asymptotically flat (AF)
metrics, which should not be reliable for non-asymptotically flat (NAF)
spacetimes. Both for magnetic \cite{9,10} and electric charges \cite{11} the
GL instability has been shown perturbatively in AF spacetimes to remain
intact.

In this paper we employ local thermodynamical stability / instability
arguments supplemented with entropy comparison to relate charged BpB and BH
in NAF spacetimes. In doing this we assume that both, the charges and the
Hawking temperatures of BpB and BH are same. Our system consists of $d-$%
dimensional Einstein-Yang-Mills (EYM) theory in a compact $p-$dimensional
brane world which admits NAF black objects. Among a large class of BH
solutions which can be used to generate a family of BpB we choose a specific
BH solution so that technically it becomes tractable. In other words, as
long as the NAF condition is assumed the freedom of alternative solutions is
always available. Our line element has the particular property that the
coefficient of the angular part, i.e. $d\Omega _{d-2}^{2}$ is a constant.
Non-Abelian gauge fields were considered as BS solutions by other
researchers \cite{12,13,14,15}. Our approach, however, differs from other
studies where we present exact non-Abelian solutions in all dimensions.
Asymptotically our solutions represent anti-de Sitter (AdS) spacetimes.
Further, by assuming an imaginary thin-shell of finite radius ($r=r_{B}$) as
boundary outside the horizon, we determine the total energy as $%
r_{B}\rightarrow \infty .$ For $d>4,$ however, the mass of our NAF black
hole diverges as $r_{B}\rightarrow \infty .$ This has been adopted as a
useful technique to define mass in NAF metrics of general relativity \cite%
{16,17}. Herein we wish to follow the same trend. This reflects the
fundamental difficulty in defining total mass for a NAF metric. \ We resort
next to compare the entropies of BpB and BH with related horizons $r_{+}$
and $r_{h},$ respectively, but with the common charge $Q$. The equality of
charges can easily be justified as the conservation of charge whereas the
relation between horizon radii of the two species is obtained from the
equality of their Hawking temperatures. From comparison of entropy
expressions we plot $r_{+}$ versus $Q^{2}$ for $5\leq d+p\leq 10$ to
identify the regions of both BpB and BH. The intersecting curve determines
naturally the transition between BpB to BH and vice versa. The figure
implies that for a given dimension $d+p=$ constant, increasing $p$ /
decreasing $d,$ favors a larger region for BH/ BpB. Assuming any relation
between the charges of BH and BpB will naturally give rise to a geometrical
constraint between two structures that undergo transition with the mass
calculation in a thin-shell formalism. Organization of the paper is as
follows.

In section II we present our exact EYM solutions in all dimensions.
Thermodynamic stability of the solution is discussed in section III. We
summarize our results in Conclusion which appears in section IV.

\section{The NAF EYM solution}

Our $(d+p)-$dimensional action with $p-$compact dimensions in EYM theory is
given by%
\begin{equation}
I=\frac{1}{16\pi G_{\left( d+p\right) }^{BpB}}\int_{0}^{L_{p}}dz_{p}...%
\int_{0}^{L_{1}}dz_{1}\int d^{d}x\sqrt{-g}\left( R-\mathcal{F}\right) 
\end{equation}%
in which 
\begin{equation}
\mathcal{F}=\mathbf{Tr}(F_{\mu \nu }^{\left( a\right) }F^{\left( a\right)
\mu \nu })
\end{equation}%
and 
\begin{equation}
\mathbf{F}^{\left( a\right) }=\mathbf{dA}^{\left( a\right) }+\frac{1}{%
2\sigma }C_{\left( b\right) \left( c\right) }^{\left( a\right) }\mathbf{A}%
^{\left( b\right) }\wedge \mathbf{A}^{\left( c\right) },
\end{equation}%
is the YM $2-$form field with structure constants $C_{\left( b\right) \left(
c\right) }^{\left( a\right) }.$ Here $R$ is the Ricci scalar, the coupling
constant $\sigma $ is expressed in terms of the YM charge and 
\begin{equation}
\mathbf{Tr}(.)=\sum\limits_{a=1}^{\frac{\left( d-1\right) (d-2)}{2}}\left(
.\right) .
\end{equation}%
Note that $G_{\left( d+p\right) }^{BpB}$ represents the $d+p-$dimensional
Newton constant while $L_{i}$ ($1\leq i\leq p$) stands for a set of compact
dimensions. We note also that $L_{i}=L,$ for all $1\leq i\leq p,$ need not
to be assumed here. For future reference we prefer to use $L_{i}\neq L_{j}$
for $i\neq j$ in this paper. As a matter of fact our results in this paper
will be valid irrespective of the compact volume. Our ultimate choice in
this study will be $\prod\limits_{i=1}^{p}L_{i}=1,$ so that if $%
L_{i}=\epsilon $ and $L_{j}=\frac{1}{\epsilon },$ for some $i$ and $j$ and $%
\epsilon \rightarrow 0$ we preserve the same volume with non-compact
translational symmetry. Our pure magnetic YM potential follows from the
higher dimensional version \cite{18,19,20,21,22} of the Wu-Yang ansatz which
is given by%
\begin{align}
\mathbf{A}^{(a)}& =\frac{Q}{r^{2}}C_{\left( i\right) \left( j\right)
}^{\left( a\right) }\ x^{i}dx^{j},\text{ \ \ }Q=\text{YM magnetic charge, \ }%
r^{2}=\overset{d-1}{\underset{i=1}{\sum }}x_{i}^{2}, \\
2& \leq j+1\leq i\leq d-1,\text{ \ and \ }1\leq a\leq \left( d-1\right)
(d-2)/2,  \notag \\
x_{1}& =r\cos \theta _{d-3}\sin \theta _{d-4}...\sin \theta _{1},\text{ }%
x_{2}=r\sin \theta _{d-3}\sin \theta _{d-4}...\sin \theta _{1},  \notag \\
\text{ }x_{3}& =r\cos \theta _{d-4}\sin \theta _{d-5}...\sin \theta _{1},%
\text{ }x_{4}=r\sin \theta _{d-4}\sin \theta _{d-5}...\sin \theta _{1}, \\
& ...  \notag \\
x_{d-2}& =r\cos \theta _{1}.  \notag
\end{align}%
The associated Lie group for the YM field is $SO(N-1)$ whereas the group of
motion over $p-$branes is the Euclidean $E^{P}.$ The overall product group
becomes therefore $E^{P}\times SO\left( N-1\right) $. Our choice for the BpB
metric ansatz is given by 
\begin{eqnarray}
ds^{2} &=&e^{-b\psi }\left( -f\left( r\right) dt^{2}+\frac{dr^{2}}{f\left(
r\right) }+h\left( r\right) ^{2}d\Omega _{\left( d-2\right) }^{2}\right) +e^{%
\frac{b\left( d-2\right) }{p}\psi }dz^{i}dz^{i}, \\
&&\left( i=1,2,...,p\right) \text{ }(b=\text{constant})  \notag
\end{eqnarray}%
in which $\psi \left( r\right) ,$ $f\left( r\right) $ and $h\left( r\right) $
are metric functions of $r$ to be found and $d\Omega _{\left( d-2\right)
}^{2}$ is the $\left( d-2\right) -$dimensional unit spherical line element.
Herein, $b$ is a constant scaling for $\psi \left( r\right) $ that plays
role in going from lower to higher dimensions or vice versa. We note that $%
b=0$, smears out the function $\psi \left( r\right) $ as well which reduces
the line element (6) to the case of AF-BpB. For this reason from the outset
we assume that the constant parameter $b$ takes values in the range $%
0<b<\infty $. In the sequel $b$ will be fixed in terms of other parameters.
Variation of the action with respect to $g_{\mu \nu }$ yields%
\begin{equation}
G_{\mu }^{\nu }=T_{\mu }^{\nu },
\end{equation}%
where 
\begin{equation}
T_{\mu }^{\nu }=2\left[ \mathbf{Tr}\left( F_{\mu \lambda }^{\left( a\right)
}F^{\left( a\right) \nu \lambda }\right) -\frac{1}{4}\mathcal{F}\delta _{\mu
}^{\nu }\right] 
\end{equation}%
or explicitly%
\begin{equation}
T_{\mu }^{\nu }=-\frac{\left( d-3\right) \left( d-2\right) Q^{2}}{2h^{4}}%
e^{2b\psi }\text{diag}\left[ 1,1,\overset{d-2-times}{\overbrace{\kappa
,\kappa ,...,\kappa }},\overset{p-times}{\overbrace{1,1,...,1}}\right] ,%
\text{ \ \ \ \ \ \ }\kappa =\frac{d-6}{d-2}
\end{equation}%
and non-zero $G_{\mu }^{\nu }$ are given by 
\begin{eqnarray}
&&%
\begin{tabular}{l}
$G_{t}^{t}=\frac{\left( d-2\right) e^{b\psi }}{2h^{2}}\left[ \frac{d+p-2}{4p}%
b^{2}fh^{2}\psi ^{\prime 2}+hh^{\prime }f^{\prime }+2fhh^{\prime \prime
}+\left( d-3\right) \left( fh^{\prime 2}-1\right) \right] $%
\end{tabular}%
, \\
&&%
\begin{tabular}{l}
$G_{r}^{r}=\frac{\left( d-2\right) e^{b\psi }}{2h^{2}}\left[ -\frac{d+p-2}{4p%
}b^{2}fh^{2}\psi ^{\prime 2}+hh^{\prime }f^{\prime }+\left( d-3\right)
\left( fh^{\prime 2}-1\right) \right] $%
\end{tabular}%
, \\
&&%
\begin{tabular}{l}
$G_{\theta _{i}}^{\theta _{i}}=\frac{e^{b\psi }}{h^{2}}\left[ \frac{d+p-2}{4p%
}\left( \frac{\left( d-2\right) }{2}b^{2}fh^{2}\psi ^{\prime 2}\right)
+\left( d-3\right) h\left( h^{\prime }f\right) ^{\prime }+\frac{\left(
d-3\right) \left( d-4\right) }{2}\left( fh^{\prime 2}-1\right) +\frac{1}{2}%
h^{2}f^{\prime \prime }\right] $%
\end{tabular}%
, \\
&&%
\begin{tabular}{l}
$G_{z_{i}}^{z_{i}}=\frac{\left( d-2\right) e^{b\psi }}{2h^{2}}\left[ \frac{%
d+p-2}{4p}\left( b^{2}fh^{2}\psi ^{\prime 2}-\frac{4b\left( f\psi ^{\prime
}\right) ^{\prime }h^{2}}{\left( d-2\right) }-4hfb\psi ^{\prime }h^{\prime
}\right) +\frac{h^{2}f^{\prime \prime }}{d-2}+2h\left( h^{\prime }f\right)
^{\prime }+\left( d-3\right) \left( fh^{\prime 2}-1\right) \right] $%
\end{tabular}%
.
\end{eqnarray}%
The YM equations also follow from the action as 
\begin{equation}
\mathbf{d}\left( ^{\star }\mathbf{F}^{\left( a\right) }\right) +\frac{1}{%
\sigma }C_{\left( b\right) \left( c\right) }^{\left( a\right) }\mathbf{A}%
^{\left( b\right) }\wedge ^{\star }\mathbf{F}^{\left( c\right) }=0
\end{equation}%
where the hodge star $^{\star }$ implies duality. By direct substitution,
one can show that YM equations are satisfied. Since this has been given
elsewhere \cite{18,19,20,21,22} it will not be repeated here. From $%
T_{t}^{t}=T_{r}^{r}$ (or $G_{t}^{t}=G_{r}^{r}$) it follows that%
\begin{equation}
\frac{\left( d+p-2\right) b^{2}}{4p}\psi ^{\prime 2}=-\frac{h^{\prime \prime
}}{h},
\end{equation}%
which after setting $h=\xi e^{\alpha \psi }$ we obtain%
\begin{equation}
\psi =\frac{\alpha }{\alpha ^{2}+\left( \frac{d+p-2}{4p}\right) b^{2}}\ln
\left( \xi _{1}r+\xi _{2}\right) .
\end{equation}%
Here $\alpha $ is a parameter to be fixed while $\xi _{1}$and $\xi _{2}$ are
two integration constants. Note also that we exclude the case $p=0$ so that
the present class of black holes doesn't admit previously known class such
as Tangherlini \cite{23} Technically we are interested in NAF solutions, we
make the choice $\xi _{1}=1$ and $\xi _{2}=0$ and consequently%
\begin{equation}
\psi =\frac{\alpha \ln r}{\alpha ^{2}+\left( \frac{d+p-2}{4p}\right) b^{2}}.
\end{equation}%
Next, we substitute $\psi $ and $h$ into the EYM BpB equations (8) and upon
the choice of $\alpha =\frac{b}{2},$ the field equations are all satisfied
with

\begin{equation}
f\left( r\right) =\Xi \left( 1-\left( \frac{r_{+}}{r}\right) ^{\frac{\left(
p+1\right) \left( d-2\right) }{d+2p-2}}\right) r^{\frac{2\left( d+p-2\right) 
}{d+2p-2}},
\end{equation}%
where 
\begin{equation}
\Xi =\frac{\left( d-3\right) \left( d+p-2\right) }{Q^{2}\left( d-2\right)
\left( p+1\right) }
\end{equation}%
and from $h=\xi e^{\alpha \psi },$ $\xi ^{2}$ is determined as 
\begin{equation}
\xi ^{2}=\frac{Q^{2}\left( d+2p-2\right) }{d+p-2}.
\end{equation}%
Here $r_{+}$ is an integration constant which represents the radius of the
event horizon and we also remark that for meaningful metric functions we
must have $d>3$. Upon rewriting the line element in the form 
\begin{equation}
ds^{2}=-f_{1}\left( r\right) dt^{2}+\frac{dr^{2}}{f_{2}\left( r\right) }%
+f_{3}\left( r\right) dz_{i}dz^{i}+f_{4}\left( r\right) d\Omega _{\left(
d-2\right) }^{2}
\end{equation}%
the metric functions take the following forms 
\begin{eqnarray}
f_{1}\left( r\right)  &=&f\left( r\right) e^{-b\psi }=\Xi \left( 1-\left( 
\frac{r_{+}}{r}\right) ^{\frac{\left( p+1\right) \left( d-2\right) }{d+2p-2}%
}\right) r^{\frac{2\left( d-2\right) }{d+2p-2}}, \\
f_{2}\left( r\right)  &=&f\left( r\right) e^{b\psi }=\Xi \left( 1-\left( 
\frac{r_{+}}{r}\right) ^{\frac{\left( p+1\right) \left( d-2\right) }{d+2p-2}%
}\right) r^{2}, \\
f_{3}\left( r\right)  &=&e^{\frac{b\left( d-2\right) }{p}\psi }=r^{\frac{%
2\left( d-2\right) }{d+2p-2}},\text{ \ \ \ }f_{4}\left( r\right) =e^{-b\psi
}h\left( r\right) ^{2}=\xi ^{2},
\end{eqnarray}%
and 
\begin{equation}
h\left( r\right) =\xi r^{\frac{p}{d+2p-2}}.
\end{equation}%
We wish now to determine the total energy for a NAF metric enclosed in an
imaginary thin-shell of radius $r=r_{B}$, where $r_{B}$ lies outside the
event horizon ($r_{B}>r_{+}$). For this purpose we consider a timelike
hypersurface $\Sigma $ defined by%
\begin{equation}
r=r_{B}.
\end{equation}%
where $r_{B}$ is the radius of the hypersurface which we call, the boundary.
By considering the constraint 
\begin{equation}
-f_{1}\left( r\right) dt^{2}+\frac{dr^{2}}{f_{2}\left( r\right) }=-d\tau ^{2}
\end{equation}%
in which $\tau $ is the proper time on the hypersurface, the line element on 
$\Sigma $ becomes%
\begin{equation}
ds_{\Sigma }^{2}=-d\tau ^{2}+f_{3}\left( r_{B}\right)
dz_{i}dz^{i}+f_{4}\left( r_{B}\right) d\Omega _{\left( d-2\right) }^{2}.
\end{equation}%
In terms of the original coordinates $x^{\gamma }=\left(
t,r,z_{1},z_{2},...,\theta _{1},\theta _{2},...\right) $ the induced metric $%
\xi ^{i}=\left( \tau ,z_{1},z_{2},...,\theta _{1},\theta _{2},...\right) $
on $\Sigma $ is given by (Latin indices run over the induced coordinates and
Greek indices run over the original manifold's coordinates)%
\begin{equation}
g_{ij}=\frac{\partial x^{\alpha }}{\partial \xi ^{i}}\frac{\partial x^{\beta
}}{\partial \xi ^{j}}g_{\alpha \beta }.
\end{equation}%
Here 
\begin{equation}
g_{ij}=\text{diag}\left( -1,f_{3}\left( r_{B}\right) ,f_{3}\left(
r_{B}\right) ,...,f_{4}\left( r_{B}\right) ,f_{4}\left( r_{B}\right)
,...\right) ,
\end{equation}%
while the extrinsic curvature is defined by 
\begin{equation}
K_{ij}=-n_{\gamma }\left( \frac{\partial ^{2}x^{\gamma }}{\partial \xi
^{i}\partial \xi ^{j}}+\Gamma _{\alpha \beta }^{\gamma }\frac{\partial
x^{\alpha }}{\partial \xi ^{i}}\frac{\partial x^{\beta }}{\partial \xi ^{j}}%
\right) _{r=r_{B}}.
\end{equation}%
It is assumed that $\Sigma $ is timelike, whose unit $\left( d+p\right) -$%
normal in $M$ is given by 
\begin{equation}
n_{\gamma }=\left( -\left\vert g^{\alpha \beta }\frac{\partial F}{\partial
x^{\alpha }}\frac{\partial F}{\partial x^{\beta }}\right\vert ^{-1/2}\frac{%
\partial F}{\partial x^{\gamma }}\right) _{r=r_{B}},
\end{equation}%
in which $F$ is the equation of the hypersurface $\Sigma ,$ i.e. 
\begin{equation}
F\left( r\right) =r-r_{B}=0.
\end{equation}%
From the Lanczos equation \cite{24,25,26} (this is the Einstein equation on
the hypersurface) the intrinsic surface stress--energy tensor, $S_{i}^{j}=$%
diag$\left( -\sigma ,p_{\theta _{1}},p_{\theta _{2}},...\right) $, is given
by 
\begin{equation}
S_{i}^{j}=-\frac{1}{8\pi G_{\left( d+p\right) }^{BpB}}\left(
K_{i}^{j}-K\delta _{i}^{j}\right) ,
\end{equation}%
in which $K$ is the trace of $K_{i}^{j}.$

Applying the above equations leads to%
\begin{equation}
g^{\alpha \beta }\frac{\partial F}{\partial x^{\alpha }}\frac{\partial F}{%
\partial x^{\beta }}=g^{rr}\left( \frac{\partial F}{\partial r}\right)
^{2}=g^{rr}=f_{2}\left( r_{B}\right)
\end{equation}%
and therefore%
\begin{equation}
n_{\gamma }=-\frac{1}{\sqrt{f_{2}\left( r_{B}\right) }}\frac{\partial F}{%
\partial x^{\gamma }}=-\frac{1}{\sqrt{f_{2}\left( r_{B}\right) }}\left(
0,1,0,0,...\right) .
\end{equation}%
This helps us to find%
\begin{gather}
K_{\tau }^{\tau }=-\frac{\sqrt{f_{2}\left( r_{B}\right) }}{2}\frac{%
f_{1}^{\prime }\left( r_{B}\right) }{f_{1}\left( r_{B}\right) } \\
K_{z_{i}}^{z_{i}}=-\frac{\sqrt{f_{2}\left( r_{B}\right) }}{2}\frac{%
f_{3}^{\prime }\left( r_{B}\right) }{f_{3}\left( r_{B}\right) } \\
K_{\theta _{i}}^{\theta _{i}}=-\frac{\sqrt{f_{2}\left( r_{B}\right) }}{2}%
\frac{f_{4}^{\prime }\left( r_{B}\right) }{f_{4}\left( r_{B}\right) },
\end{gather}%
and consequently%
\begin{equation}
K=K_{i}^{i}=-\frac{\sqrt{f_{2}\left( r_{B}\right) }}{2}\frac{f_{1}^{\prime
}\left( r_{B}\right) }{f_{1}\left( r_{B}\right) }-\frac{p\sqrt{f_{2}\left(
r_{B}\right) }}{2}\frac{f_{3}^{\prime }\left( r_{B}\right) }{f_{3}\left(
r_{B}\right) }-\frac{\left( d-2\right) \sqrt{f_{2}\left( r_{B}\right) }}{2}%
\frac{f_{4}^{\prime }\left( r_{B}\right) }{f_{4}\left( r_{B}\right) }.
\end{equation}%
The energy density $\sigma $ on the boundary and pressure $p_{i}$ are as
follow%
\begin{gather}
\sigma =-S_{\tau }^{\tau }=\frac{1}{8\pi G_{\left( d+p\right) }^{BpB}}\left( 
\frac{p\sqrt{f_{2}\left( r_{B}\right) }}{2}\frac{f_{3}^{\prime }\left(
r_{B}\right) }{f_{3}\left( r_{B}\right) }+\frac{\left( d-2\right) \sqrt{%
f_{2}\left( r_{B}\right) }}{2}\frac{f_{4}^{\prime }\left( r_{B}\right) }{%
f_{4}\left( r_{B}\right) }\right) , \\
p_{\theta _{i}}=\frac{1}{8\pi G_{\left( d+p\right) }^{BpB}}\left( -\frac{%
\sqrt{f_{2}\left( r_{B}\right) }}{2}\frac{f_{1}^{\prime }\left( r_{B}\right) 
}{f_{1}\left( r_{B}\right) }-\frac{p\sqrt{f_{2}\left( r_{B}\right) }}{2}%
\frac{f_{3}^{\prime }\left( r_{B}\right) }{f_{3}\left( r_{B}\right) }-\frac{%
\left( d-3\right) \sqrt{f_{2}\left( r_{B}\right) }}{2}\frac{f_{4}^{\prime
}\left( r_{B}\right) }{f_{4}\left( r_{B}\right) }\right) , \\
p_{z_{i}}=\frac{1}{8\pi G_{\left( d+p\right) }^{BpB}}\left( -\frac{\sqrt{%
f_{2}\left( r_{B}\right) }}{2}\frac{f_{1}^{\prime }\left( r_{B}\right) }{%
f_{1}\left( r_{B}\right) }-\frac{\left( p-1\right) \sqrt{f_{2}\left(
r_{B}\right) }}{2}\frac{f_{3}^{\prime }\left( r_{B}\right) }{f_{3}\left(
r_{B}\right) }-\frac{\left( d-2\right) \sqrt{f_{2}\left( r_{B}\right) }}{2}%
\frac{f_{4}^{\prime }\left( r_{B}\right) }{f_{4}\left( r_{B}\right) }\right)
.
\end{gather}%
In order to find the total energy (=Mass), we use the following integral 
\begin{equation}
\Omega =\int \sqrt{-g}\left( \rho +p_{r}\right) d^{d+p}x
\end{equation}%
in which $p_{r}$ is the radial pressure. For the boundary surface we have $%
p_{r}=0$ and $\rho =\sigma \delta \left( r-r_{B}\right) $ where $\delta
\left( r-r_{B}\right) $ stands for the Dirac delta function. A simple
calculation results in (keeping in mind that each $z_{i}$ is compact and
contributes trivially)%
\begin{eqnarray}
\Omega
&=&\int\limits_{0}^{L_{p}}...\int\limits_{0}^{L_{1}}\int\limits_{0}^{2\pi
}\int\limits_{0}^{\pi }...\int\limits_{0}^{\pi }\int\limits_{0}^{\infty }%
\sqrt{-g}\sigma \delta \left( r-r_{B}\right) drd\theta _{1}...d\theta
_{d-2}dz_{1}...dz_{p}= \\
&&\frac{\pi ^{\frac{d-1}{2}}\prod\limits_{k=1}^{p}L_{k}}{8\pi \Gamma \left( 
\frac{d-1}{2}\right) G_{\left( d+p\right) }^{BpB}}\sqrt{f_{1}\left(
r_{B}\right) f_{3}\left( r_{B}\right) ^{p}f_{4}\left( r_{B}\right) ^{d-2}}%
\left( p\frac{f_{3}^{\prime }\left( r_{B}\right) }{f_{3}\left( r_{B}\right) }%
+\left( d-2\right) \frac{f_{4}^{\prime }\left( r_{B}\right) }{f_{4}\left(
r_{B}\right) }\right) .  \notag
\end{eqnarray}%
In terms of our metric functions one finds%
\begin{equation}
\Omega =\frac{\pi ^{\frac{d-1}{2}}\xi ^{d-2}\sqrt{\Xi }\prod%
\limits_{k=1}^{p}L_{k}}{4\pi \Gamma \left( \frac{d-1}{2}\right) G_{\left(
d+p\right) }^{BpB}}\frac{\left( d-2\right) p}{d+2p-2}\frac{1}{r_{B}}\left(
1-\left( \frac{r_{+}}{r_{B}}\right) ^{\frac{\left( p+1\right) \left(
d-2\right) }{d+2p-2}}\right) ^{1/2}r_{B}^{\frac{\left( d-2\right) \left(
p+1\right) }{d+2p-2}},
\end{equation}%
which is in our case the total energy of the spacetime stored inside $%
r=r_{B}.$ One may call it the mass of the solution. In the case of $\left(
4+p\right) -$dimensions we find%
\begin{equation}
M=\Omega =\frac{1}{G_{\left( d+p\right) }^{BpB}}\sqrt{\frac{p^{2}Q^{2}}{%
2\left( d+2\right) \left( p+1\right) }}\prod\limits_{k=1}^{p}L_{k}\left( 1-%
\frac{r_{+}}{r_{B}}\right) ^{1/2},
\end{equation}%
which in the limit of $r_{B}\rightarrow \infty $ becomes%
\begin{equation}
M=\frac{1}{G_{\left( d+p\right) }^{BpB}}\sqrt{\frac{p^{2}Q^{2}}{2\left(
d+2\right) \left( p+1\right) }}\prod\limits_{k=1}^{p}L_{k},
\end{equation}%
and is finite. We admit, however, that for $d>4$ (with $p=$arbitrary) the
mass expression (47) diverges when $r_{B}\rightarrow \infty .$ In order to
understand the physical implication of this class of NAF solutions we
investigate their asymptotic behaviors for $r\gg r_{+}$. For this purpose we
make the transformation 
\begin{eqnarray}
r &\rightarrow &r^{-k}  \notag \\
\text{(}k &=&\frac{d+2p-2}{d-2}\text{),}
\end{eqnarray}%
followed by the scalings%
\begin{eqnarray}
t &\rightarrow &\left( \frac{\Xi }{k}\right) t,  \notag \\
z_{i} &\rightarrow &\left( \frac{\sqrt{\Xi }}{k}\right) z_{i}
\end{eqnarray}%
to yield for $r\gg r_{+}$%
\begin{equation}
ds^{2}\simeq \frac{1}{r^{2}}\left( -dt^{2}+dr^{2}+dz_{i}dz^{i}\right) +\frac{%
d-3}{k\left( p+1\right) }d\Omega _{\left( d-2\right) }^{2}.
\end{equation}%
This is the geometry of the $(p+2)-$dimensional anti-de Sitter (AdS)
spacetime times the $(d-2)-$dimensional sphere (i.e. $AdS_{p+2}\times
S^{d-2} $). In analogy, near the horizon, i.e., $r=r_{+}+x$ with $x>0,$ ($%
x^{2}\approx 0$) we have our line element 
\begin{equation}
ds^{2}\simeq -C_{0}xdt^{2}+\frac{dx^{2}}{C_{1}x}+C_{2}dz_{i}dz^{i}+\xi
^{2}d\Omega _{\left( d-2\right) }^{2}
\end{equation}%
for appropriate constants $C_{i}.$ Applying now 
\begin{equation}
x=\frac{1}{4}C_{1}y^{2}
\end{equation}%
and rescaling the time coordinate casts the metric into%
\begin{equation}
ds^{2}\simeq -y^{2}d\bar{t}^{2}+dy^{2}+C_{2}dz_{i}dz^{i}+\xi ^{2}d\Omega
_{\left( d-2\right) }^{2}
\end{equation}%
which is the accelerated (Rindler) frame at the ($\bar{t},y$) (or $t,r$)
sector. This is a product space of $2-$dimensional flat space with $p-$torus
and $\left( d-2\right) -$dimensional sphere (i.e., $M^{2}\times T^{p}\times
S^{d-2}$). From the general solution (23-26), for certain values of ($d,p$)
we find the Ricci scalar ($R$) and Kretschmann scalar ($K$) as follow%
\begin{equation}
d=4,\text{ \ }p=1,\text{ \ }R=\frac{3}{4Q^{2}},\text{ \ }K=\frac{171}{64Q^{4}%
},
\end{equation}%
\begin{equation}
d=4,\text{ \ }p=2,\text{ \ }R=\frac{4}{9Q^{2}},\text{ \ }K=\frac{16}{243Q^{4}%
}\left( 29+\left( \frac{r_{+}}{r}\right) ^{2}\right) ,
\end{equation}%
\begin{equation}
d=5,\ p=1,\ R=\frac{3}{4Q^{2}},\ K=\frac{6528}{625Q^{4}},\text{ \ \ }...%
\text{ .}
\end{equation}%
One characteristic feature of this class of solutions is that whenever $p=1$%
, irrespective of $d$, we have a regular solution while for $p>1$ it is
singular at $r=0$.

The simplest member in this class of solutions is given by the choice $%
d=4,p=1$. This yields the line element of a black string \cite{27} 
\begin{equation}
ds^{2}=-\frac{3\left( r-r_{+}\right) }{4Q^{2}}dt^{2}+\frac{4Q^{2}}{3r\left(
r-r_{+}\right) }dr^{2}+rdz^{2}+\frac{4Q^{2}}{3}d\Omega _{2}^{2}
\end{equation}%
which is non-singular and manifestly NAF. The scalar curvature is $R=\frac{3%
}{8Q^{2}}$, and the Kretschmann scalar is $K=\frac{171}{64Q^{4}}$. From our
foregoing argument this asymptotes (for $r\gg r_{+}$) to the spacetime $%
adS_{3}\times S^{2}$ while for $r\approx r_{+}$ it is $M^{2}\times
S^{1}\times S^{2}.$ We note that our solutions are generically regular for $%
p=1$, and singular at $r=0$ for $p>1$. The singularity shows itself in the
Kretschmann scalar while other scalars are all regular. Further, the mass
for such a black string has already been defined in (47), which turns out to
be finite.

Our aim next, is to compare the entropy of $\left( d+p\right) -$dimensional
NAF-EYM BpB with the entropy of the $\left( d+p\right) -$dimensional
NAF-EYMBH whose metric is given by%
\begin{equation}
ds^{2}=-f\left( r\right) dt^{2}+\frac{dr^{2}}{f\left( r\right) }%
+h^{2}d\Omega _{\left( d+p-2\right) }^{2}
\end{equation}%
where $h$ is a constant to be fixed. The corresponding action for the $d+p-$%
dimensional EYM theory is also given by 
\begin{equation}
I_{BH}^{\left( d+p\right) }=\frac{1}{16\pi G_{\left( d+p\right) }^{BH}}\int
d^{d+p}x\sqrt{-g}\left( R-\mathcal{F}\right) 
\end{equation}%
in which $\mathcal{F}$ is the YM invariant (2). The YM equations are
satisfied with Einstein tensor components as%
\begin{equation}
G_{\mu }^{\nu }=\left[ -\frac{\left( d+p-3\right) \left( d+p-2\right) }{%
2h^{2}},-\frac{\left( d+p-3\right) \left( d+p-2\right) }{2h^{2}},\frac{%
-\left( d+p-3\right) \left( d+p-4\right) +f^{^{\prime \prime }}h^{2}}{2h^{2}}%
,...\right] .
\end{equation}%
The energy-momentum tensor components follow from (9) with $b=0$ and $%
d\rightarrow d+p,$ i.e., 
\begin{equation}
T_{\mu }^{\nu }=-\frac{\left( d+p-3\right) \left( d+p-2\right) Q_{BH}^{2}}{%
2h^{4}}\text{diag}\left[ 1,1,\overset{d+p-2-times}{\overbrace{\kappa ,\kappa
,...,\kappa }}\right] ,\text{ \ \ \ \ \ \ }\kappa =\frac{d+p-6}{d+p-2}.
\end{equation}%
The Einstein's equations (8) imply now that%
\begin{equation}
h^{2}=Q_{BH}^{2}
\end{equation}%
with%
\begin{equation}
f\left( r\right) =\frac{d+p-3}{Q_{BH}^{2}}r^{2}+C_{1}r+C_{2},
\end{equation}%
where $Q_{BH}$ stands for the charge of the BH. Here $C_{1}$ and $C_{2}$ are
two integration constants which for technical reason we set $C_{1}=0$ and $%
C_{2}=-\frac{d+p-3}{Q_{BH}^{2}}r_{h}^{2}$ to cast $f\left( r\right) $ into
the form%
\begin{equation}
f=\frac{d+p-3}{Q_{BH}^{2}}r^{2}\left( 1-\left( \frac{r_{h}}{r}\right)
^{2}\right) 
\end{equation}%
where $r_{h}$ indicates the horizon of the black hole. The entropy of
NAF-EYMBH is given by%
\begin{equation}
S_{BH}=\frac{A_{H}}{4G_{\left( d+p\right) }^{BH}}=\frac{8\pi ^{\frac{d+p+1}{2%
}}}{\Gamma \left( \frac{d+p-1}{2}\right) }\left( Q_{BH}^{2}\right) ^{\frac{%
\left( d+p-2\right) }{2}},
\end{equation}%
and%
\begin{equation}
T_{BH}=\left. \frac{f^{\prime }}{4\pi }\right\vert _{r=r_{h}}=\frac{2\left(
d+p-3\right) }{4\pi Q_{BH}^{2}}r_{h}
\end{equation}%
in which for $(d+p)-$dimensional BH we have used $16\pi G_{\left( d+p\right)
}^{BH}=1$ (i.e. the volume of the extra space due to branes is chosen as $%
\prod\limits_{k=1}^{p}L_{k}=1$). Note also that the ansatz (60) with $%
h=const.$ dashes hopes to admit Tangherlini type black holes \cite{23}.

\section{Local Thermodynamical Stability}

The entropy of the BpB metric (22) is defined by 
\begin{equation}
S_{BpB}=\frac{A_{H}}{4G_{\left( d+p\right) }^{BpB}},
\end{equation}%
in which $G_{\left( d+p\right) }^{BpB}=G_{\left( d\right)
}\prod\limits_{k=1}^{p}L_{k}$ while 
\begin{equation}
A_{H}=\frac{2\pi ^{\frac{d-1}{2}}\prod\limits_{k=1}^{p}L_{k}}{\Gamma \left( 
\frac{d-1}{2}\right) }f_{4}\left( r_{+}\right) ^{\frac{d-2}{2}}f_{3}\left(
r_{+}\right) ^{\frac{p}{2}}.
\end{equation}%
Here we set $16\pi G_{\left( d\right) }=1$ and therefore%
\begin{equation}
S_{BpB}=\frac{8\pi ^{\frac{d+1}{2}}}{\Gamma \left( \frac{d-1}{2}\right) }%
f_{4}\left( r_{+}\right) ^{\frac{d-2}{2}}f_{3}\left( r_{+}\right) ^{\frac{p}{%
2}}.
\end{equation}%
which, upon substitution from above, implies%
\begin{equation}
S_{BpB}=\frac{8\pi ^{\frac{d+1}{2}}}{\Gamma \left( \frac{d-1}{2}\right) }%
\left( \frac{Q^{2}\left( d+2p-2\right) }{d+p-2}\right) ^{\frac{d-2}{2}%
}r_{+}^{\frac{p\left( d-2\right) }{d+2p-2}}.
\end{equation}%
Now, the Hawking temperature $T_{BpB}$ \cite{28} and specific heat capacity $%
C_{Q}$ of the BpB are given by 
\begin{equation}
T_{BpB}=\left. \frac{\sqrt{f_{1}^{\prime }f_{2}^{\prime }}}{4\pi }%
\right\vert _{r=r_{+}}=\frac{\left( d+p-2\right) \left( d-3\right) }{4\pi
\left( d+2p-2\right) Q^{2}}r_{+}^{\frac{d-2}{d+2p-2}}
\end{equation}%
and%
\begin{equation}
C_{Q}=T_{H}\left( \frac{\partial S_{BpB}}{\partial T_{H}}\right) _{Q}=\frac{%
8\pi ^{\frac{d+1}{2}}p}{\Gamma \left( \frac{d-1}{2}\right) }\left( \frac{%
Q^{2}\left( d+2p-2\right) }{d+p-2}\right) ^{\frac{d-2}{2}}r_{+}^{\frac{%
p\left( d-2\right) }{d+2p-2}}.
\end{equation}%
It is observed that $S_{BpB},$ $T_{BpB}$ and $C_{Q}$ are regular and
positive in all dimensions $d>3$ which is of our case of study.

Our final argument is to define the micro-canonical equilibrium condition
for the EYM-BpB as $S_{BpB}\geq S_{BH}$ i.e.,%
\begin{equation}
\frac{\Gamma \left( \frac{d+p-1}{2}\right) }{\Gamma \left( \frac{d-1}{2}%
\right) }\left( \frac{d+2p-2}{d+p-2}\right) ^{\frac{d-2}{2}}\left(
Q_{BpB}^{2}\right) ^{\frac{d-2}{2}}r_{+}^{\frac{p\left( d-2\right) }{d+2p-2}%
}\geq \pi ^{\frac{p}{2}}\left( Q_{BH}^{2}\right) ^{\frac{\left( d+p-2\right) 
}{2}}.
\end{equation}%
By assuming now that $Q_{BH}^{2}=Q_{BpB}^{2}=Q^{2}$ as a requirement of
charge conservation in case there is a transition, leads us to the
corresponding condition defined by $S_{BpB}\geq S_{BH}$. This implies that%
\begin{equation}
r_{+}\geq \left( \frac{\Gamma \left( \frac{d-1}{2}\right) \left( \pi
Q^{2}\right) ^{\frac{p}{2}}}{\Gamma \left( \frac{d+p-1}{2}\right) }\left( 
\frac{d+p-2}{d+2p-2}\right) ^{\frac{d-2}{2}}\right) ^{\frac{d+2p-2}{p\left(
d-2\right) }}.
\end{equation}%
Fig. 1 displays the curves of equality conditions for $d+p=10$,
respectively. Above / below each curve given, BpB / BH are the corresponding
favored regions. Each curve represents the critical boundary between a BpB
and a BH. Comparison of entropies (from Eq. 75) suggests that the left (or
up) of each curve represents a BpB while the right (or down) of each curve
favors the BH state. For a constant $r_{+}$ it is observed that increment in
charge transforms a BpB into a BH. Conversely, for a fixed charge,
increasing the horizon radius $r_{+}$ goes toward BpB from the BH state.

\begin{figure}[h]
\includegraphics[width=150mm,scale=0.7]{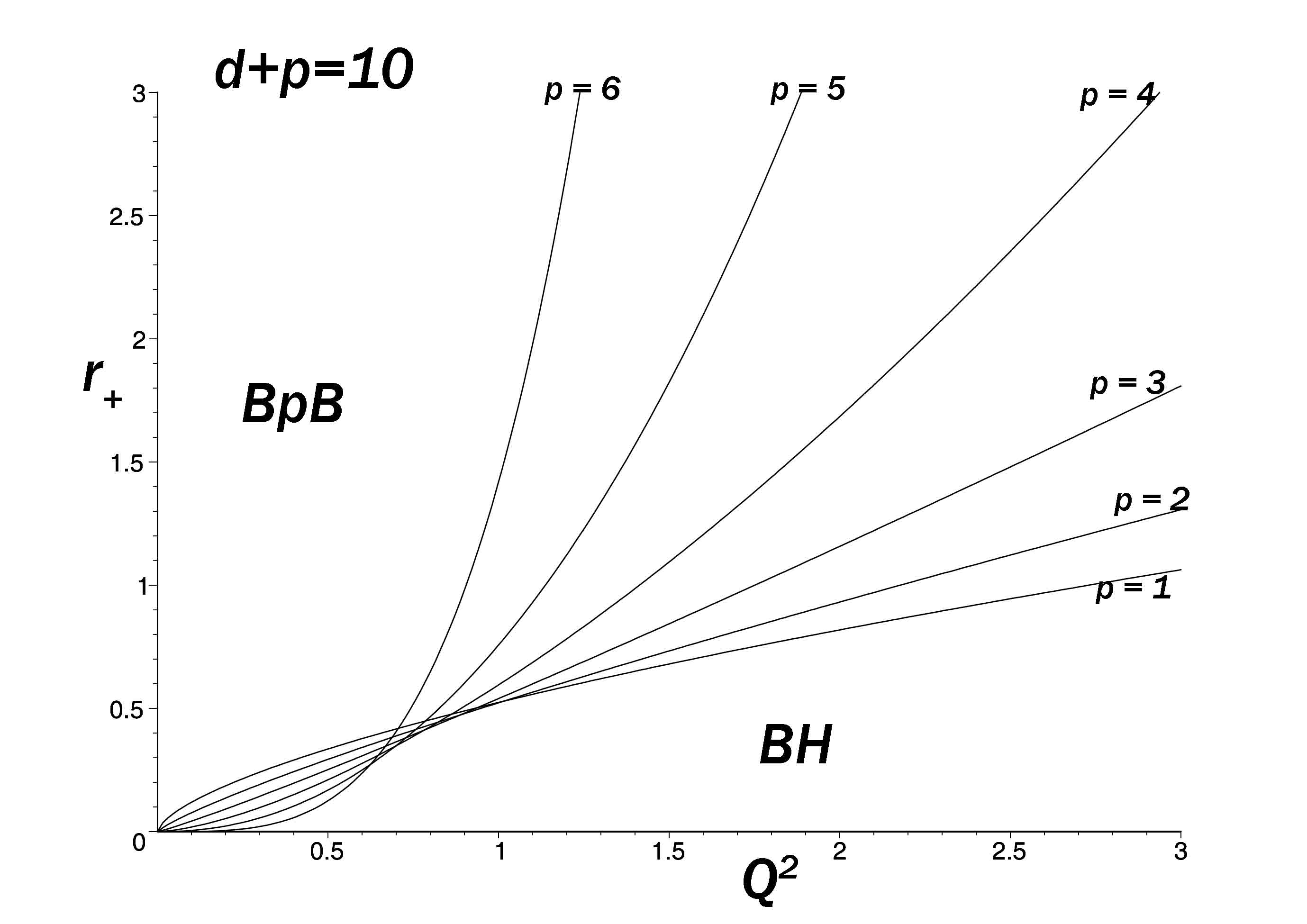}
\caption{As a result of the entropy argument for the particular case $d+p=10$%
, we obtain this informative plot of BpB effective horizon radius $r_{+}$
versus common charge square $Q^{2}$. The horizon radius of BH, $r_{h}$, is
connected to the horizon radius of BpB, $r_{+}$, through Eq. (77).}
\label{fig: 1}
\end{figure}
Let's add that, the condition of equal temperature of the BpB and BH at the
transition time yields a relation between the horizons of the two objects.
This can be seen by equating (68) and (73) to find%
\begin{equation}
r_{h}=\frac{\left( d+p-2\right) \left( d-3\right) }{2\left( d+p-3\right)
\left( d+2p-2\right) }r_{+}^{\frac{d-2}{d+2p-2}}.
\end{equation}%
Therefore in Fig. 1, in the BH region, one has to compute the target black
hole horizon radius by applying Eq. (77).

\section{Conclusion}

It is well-known that in asymptotically flat black holes stability lies at
the heart of uniqueness which makes the Birkhoff theorem. By employing the
thin-shell boundary condition we determine the mass available inside a shell
of radius $r_{B}>r_{+}.$ In NAF-EYM theory we obtained for a particular
class of solutions a critical boundary curve that separates BpB from BH
bearing connected horizon radii but common charges $Q$. We argue that our
stability treatment doesn't depend on the particular solution but is more
general. The NAF character manifests itself asymptotically as an AdS
spacetime with a suitable effective cosmological constant. The critical
curve arises from entropy comparison for the two types of black objects. It
reflects the relative weight of dimensions $p$ and $d$ for each given case $%
d+p\geqslant 5$. In particular, we plotted $r_{+}$ versus $Q^{2}$ to
represent the cases of $d+p=10$. The entropy comparison remains still a
reliable test to check possible transitions from BpB to BH and vice versa.
In this regard, we admit that our method applies only for charged black
objects of special kind which doesn't work for the neutral ones. We comment
finally that in a recent study attention is drawn to the possibility of
decay from a $5-$dimensional string into a set of $4-$dimensional naked
singularities \cite{29}. Transition into naked singularities has not been
considered in the present work.

\bigskip 

\end{document}